\documentclass[aps,nofootinbib,showpacs,preprintnumbers,amsmath,amssymb]{revtex4}
\usepackage{epsfig}

\begin{document}
\preprint{USM-TH-134; hep-ph/0211226 v3}

\title{Infrared renormalons and analyticity structure in pQCD\footnote{
hep-ph/0211226 v3: improved version - the paragraph containing
Eqs.(18) and (19) is new, as well as the next paragraph; the title modified; 
some references added; version to appear in Phys. Rev. D}
}

\author{Gorazd Cveti\v{c}}
  \email{gorazd.cvetic@fis.utfsm.cl}
\affiliation{Dept.~of Physics, Universidad T\'ecnica
Federico Santa Mar\'{\i}a, Valpara\'{\i}so, Chile}

\date{\today}

\begin{abstract}
Relation between the infrared renormalons, the Borel 
resummation prescriptions, and the analyticity structure
of Green functions in perturbative QCD (pQCD)
is investigated. A specific recently suggested Borel
resummation prescription resulted in the Principal Value 
and an additional power-suppressed correction that is 
consistent with the Operator Product Expansion. 
Arguments requiring the finiteness of the result for any 
power coefficient of the leading infrared renormalon, and 
the consistency in the case of the absence of that renormalon, 
require that this prescription be modified. The apparently 
most natural modification leads to the result represented
by the Principal Value. The analytic structure of the
amplitude in the complex coupling plane, obtained in this way,
is consistent with that obtained in the literature
by other methods. 
\end{abstract}

\maketitle

%\section{Introduction}
%\label{s1}
Green functions in QCD are quantities which, in general,
are known to possess renormalons, i.e., singularities
of the Borel transform on the real axis \cite{Beneke:1999ui}. 
They appear as a consequence of the specific asymptotic behavior 
of the high-order perturbation coefficients. The
singularities on the positive axis, called infrared
(IR) renormalons, represent an
obstacle to the Borel integration and lead to
the well-known renormalon-induced ambiguity for the
observable. The most widely used prescription
for fixing this ambiguity has been to perform the
Borel integration parallell to the positive real axis
and taking the real part of the resulting integral
(Principal Value; see, e.g., 
Refs.~\cite{Khuri:bf,Cvetic:2001sn}). 
This prescription in pQCD was shown to be favored by or 
to be consistent with analyticity requirements in the
momentum plane \cite{Caprini:1999ma} and in the coupling parameter
plane \cite{Caprini:2001mn}. Further,
this prescription is attractive due to its mathematical simplicity.
If the QCD quantity under consideration allows to be
presented by the Operator Product Expansion (OPE),
then any additional, nonperturbative, terms are expected to have the 
power-suppressed form of the higher-twist terms.
The perturbative QCD (pQCD), involving the quark and gluon
degrees of freedom, is usually expected to be unable to 
predict the strength of such terms.\footnote{The IR renormalons predict
the energy dependence of such terms.}

Recently, a specific prescription, based on the IR renormalon 
considerations, has been proposed \cite{Lee:2001ws}
to fix the strength of such higher-twist terms,
and this method gave encouraging numerical results
in the case of the resummation of the Gross--Lewellyn-Smith 
sum rule and of the heavy-quark potential \cite{Contreras:2002kf}.
The main observation was that the IR renormalon induces in the
Borel-integrated quantity a nonphysical cut along the positive 
axis in the complex plane of the coupling parameter $z$, 
and that this cut structure can be naturally eliminated by 
subtracting a cut function proportional to $(-z)^{\nu}$ where 
$\nu$ is related with the power coefficient of the renormalon
singularity. Further, the energy dependence of the subtraction 
term, when the coupling parameter $z$ is positive,
is consistent with the predictions of the
OPE for the corresponding higher-twist term.
For positive $z$, the result is the Principal Value 
of the Borel integration minus the aforementioned term.
A somewhat speculative interpretation of this result
suggests that in this way the dominant part of the genuine
nonperturbative higher-twist effect is obtained
(the correction term to the Principal Value), 
although the method is based on perturbative 
(pQCD $+$ renormalons) knowledge
only. A more conservative rephrasing of this would be that
this result represents ``the most that we can get''
out of pQCD, i.e., the natural basis to which
one should eventually add other contributions
to the aforementioned higher-twist term; such genuine
nonperturbative contributions would involve 
the vacuum expectation values of the higher-twist
operators appearing in the OPE.

In the present work, the aforementioned method is
scrutinized. As a result, more support is given
to the second of the two mentioned interpretations of the
pQCD renormalon resummation results. Even more so,
the results suggest that the natural resummed 
perturbative contribution is just the Principal Value
of the Borel integral, without the aforementioned
higher-twist term.
As a by-product, some insights into the role of the
IR renormalons in the analyticity structure of the
pQCD amplitudes in the coupling plane are obtained.

Let us consider a Euclidean QCD amplitude $\Delta[a(Q)]$ where
the quark mass effects are neglected. Therefore, it 
can be regarded as depending on the energy 
$Q \equiv \sqrt{- q^2}$ of the corresponding process
only via the QCD coupling parameter
$a(Q) \equiv \alpha_s(Q;{\overline {\rm MS}})/\pi$
whose running is determined by the (${\overline {\rm MS}}$)
renormalization group equation
\begin{equation}
\frac{ \partial a(\mu)}{\partial \ln \mu^2}
= - \beta_0 ~a^2 ~(1 + c_1 ~a + c_2 ~a^2 + c_3 ~a^3 + \cdots) \ .
\label{RGE}
\end{equation}
For simplicity of argument, we will consider only the
effects of the leading IR renormalon and will neglect 
the subleading IR renormalons. We will also assume that
such an amplitude can be described by the OPE.

\begin{figure}[htb]
\begin{center}
 \epsfig{file=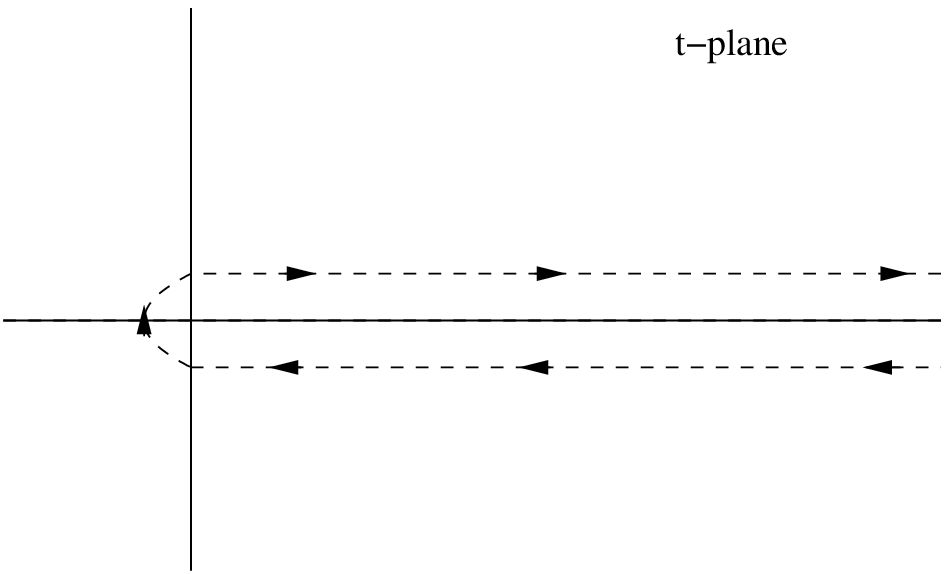, width=10.cm}
\end{center}
\caption{\footnotesize The path ${\cal C}_t$ in the integrals
(\ref{ImDBI}) and (\ref{Hankel}). 
}
\label{path}
\end{figure}

First, let us review the method of Ref.~\cite{Lee:2001ws} 
in detail. The presentation of the method is here somewhat different.
The Borel transform $B(b)$ of $\Delta[a(Q)]$
has a singularity at $b \geq n$ ($n=1$, or $2$, or $3$, ...)
of the form
\begin{eqnarray}
B(b) & = & \frac{C}{ (1 - b/n)^{1 + \nu} } \left[
1 + \kappa_1 ~(1 - b/n) + \kappa_2 ~(1 - b/n)^2 \cdots \right] 
+ {\rm analytic \ part} \ .
\label{IR1}
\end{eqnarray}
Here $\nu \!=\!(n c_1\!-\!\gamma_{2n})/\beta_0$,
where $\gamma_{2n}$ is the one-loop coefficient of the 
anomalous dimension of the
corresponding higher-twist operator (with dimension $d\!=\!2n$) 
in the OPE for $\Delta[a(Q)]$. 
In the following, we will ignore the contributions of
the terms with coefficients $\kappa_j$ ($j \geq 1$),
since their inclusion is straightforward ($\nu \mapsto \nu - j$)
and does not affect the conclusions.
For definiteness, the renormalization
scale $\mu$ is taken to be $\mu\!=\!Q$, and the
renormalization scheme is also regarded as fixed, e.g., 
${\overline {\rm MS}}$. We will regard the coupling to
be complex in general, $z\!\equiv\!\beta_0 a(Q)/n\!=\!|z| \exp(i \phi)$,
corresponding to complex momenta $Q$.
Then the Borel integral $\Delta_{\rm BI}(z)$ can be defined as 
\begin{equation}
\Delta_{\rm BI}(z) = \frac{1}{\beta_0}~
\int_{0}^{+ \infty \exp(i \phi)}
db~\exp \left(- \frac{b}{n z} \right) B(b) \ , \qquad
(z = |z|e^{i \phi}, b = |b| e^{i \phi}) \ .
\label{Bintegraldef}
\end{equation}
For $z$ near the positive real axis
$z = |z| \pm i \varepsilon$ [$|z|\!=\!\beta_0 a(Q)/n > 0$], 
it is straightforward to show from (\ref{Bintegraldef}), 
with the help of the Cauchy theorem, the following formula: 
\begin{eqnarray}
\Delta_{\rm BI}(z) {\big |}_{z = |z| \pm i \varepsilon}
&=& \frac{1}{\beta_0}
~\int_{\pm i \varepsilon}^{+\infty \pm i \varepsilon} 
~db~\exp \left(- \frac{b}{n |z|} \right) B(b) 
\qquad (\varepsilon \to +0) \ .
\label{Bintegral}
\end{eqnarray}
The real part of this is the Principal Value.
As a consequence of the IR singularity (\ref{IR1}) 
at $b \geq n$, the integral (\ref{Bintegraldef})
has a discontinuity (cut)
at the positive real axis $z \geq 0$. Namely,
the quantity (\ref{Bintegral}) is not real and its discontinuity 
shows up in its imaginary part. This can be seen by introducing
in the integrand of Eq.~(\ref{Bintegral}) the new complex
integration variable $t$ via $b = n(1 + |z| t)$
\begin{eqnarray}
{\rm Im} \Delta_{\rm BI}(z=|z|\!\pm\!i \varepsilon) &=&
\pm \frac{1}{2 i \beta_0} 
\int_{{\cal C}_t}
db \exp \left(- \frac{b}{n |z|} \right)
\frac{C}{(1\!-\!b/n)^{1\!+\!\nu}} 
= \mp C \frac{n}{\beta_0} ~e^{-1/|z|}
~\Gamma(-\nu)~\sin( \pi \nu) ~|z|^{-\nu} \ .
\label{ImDBI}
\end{eqnarray}
The (Hankel) contour ${\cal C}_t$ is depicted in Fig.~\ref{path},
and the expression (\ref{ImDBI}) is obtained from the
known Hankel contour form of the Gamma function
(see, for example, \cite{GR})
\begin{equation}
 \Gamma(s) \sin(\pi s) = \frac{\pi}{\Gamma(1 - s)} =
+ \frac{1}{2 i} \int_{{\cal C}_t}
dt e^{-t} (-t)^{-1 + s} \qquad (|s| < \infty) \ ,
\label{Hankel}
\end{equation}
in the special cases $s = -\nu, -\nu + 1, \ldots$. 
\noindent
Because $(-|z|\!\mp\!i \varepsilon) = |z| \exp( \mp i \pi)$ and thus
$(-|z|\!\mp\!i \varepsilon)^{-\nu}\!=\!
|z|^{-\nu} [ \cos(\pi \nu)\!\pm\!i \sin(\pi \nu)]$,
Eq.~(\ref{ImDBI}) leads to
\begin{equation}
\Delta_{\rm BI}(z) =  - C \frac{n}{\beta_0} ~e^{-1/z}
~\Gamma(-\nu)~(- z)^{-\nu}
+ {\tilde \Delta}_{\rm BI}(z),
\label{DBI}
\end{equation} 
where ${\tilde \Delta}_{\rm BI}(z)$ is a function without
cuts in the complex $z$-plane since the first
expression on the right-hand side absorbs
the renormalon-induced cut (\ref{ImDBI}) at $z \geq 0$.
The first expression, when $z\!=\!|z|\!\pm\!i \varepsilon =
\beta_0 a(Q)/n\!\pm\!i \varepsilon$ is at the real positive axis, 
has the $Q$-dependence 
$Q^{-2n} \alpha_s(Q)^{\gamma_{2n}/\beta_0} [ 1\!+\!{\cal O}(\alpha_s) ]$, 
the same as the corresponding power-suppressed (higher-twist) term 
$\langle {\cal O}_{2n} \rangle^{(Q)}/(Q^2)^n$ in the OPE. 
The imaginary part of this
expression, i.e., expression (\ref{ImDBI}), must be identified 
as the imaginary part of the
contribution from the leading IR renormalon (\ref{IR1}).
The central assumption of the method is 
that the full first expression on 
the right-hand side of (\ref{DBI}), which
contains the full nonphysical cut (\ref{ImDBI}) 
along $z >0$ and no cut along $z < 0$,
represents the nonphysical cut-function which is to be eliminated
\begin{equation}
\Delta_{\rm (cut1)}(z) = 
- C \frac{n}{\beta_0} ~e^{-1/z} ~\Gamma(-\nu)~(- z)^{-\nu} 
= C \frac{n}{\beta_0} ~e^{-1/z} 
~\frac{\pi}{\Gamma(1\!+\!\nu) \sin(\pi \nu)}~(- z)^{-\nu} 
\ .
\label{cut1}
\end{equation}
This same cut contribution, for $z\!=\!|z|\!\pm\!i \varepsilon$ at the
positive real axis, can be obtained also by Borel-integrating
the nonanalytic part of the Borel transform (\ref{IR1}) 
along its cut $b > n$ above or below
the real axis in analogy with the full Borel-integrated
quantity (\ref{Bintegral}). To show this, we use the new
real integration variable $t$ such that $b\!=\!n(1\!+\!z t)$,
and therefore $b > n \pm i \varepsilon$ corresponds to $t > 0$
\begin{eqnarray}
\Delta_{\rm BI}^{({\rm nonan.})}(z\!=\!|z|\!\pm\!i \varepsilon; b > n)
&=&
\frac{1}{\beta_0}
~\int_{n \pm i \varepsilon}^{\infty \pm i \varepsilon} 
~db~\exp \left(- \frac{b}{n z} \right) B^{(\rm nonan.)}(b)
\label{cut1p}
\\
& = & + C \frac{n}{\beta_0}~z~e^{-1/z}~\int_0^{+\infty}
dt~e^{-t}~(- z)^{-1 - \nu}~t^{-1 - \nu}
\label{cut1pp} 
\\
&=&
- C \frac{n}{\beta_0} ~e^{-1/z}
~\Gamma(-\nu)~(- z)^{-\nu} \ .
\label{cut1ppp}
\end{eqnarray}
For $z$ away from the real positive axis, the analytic continuation
of this expression in $z$ keeps its form (\ref{cut1ppp}) unchanged,
i.e., precisely the cut-function (\ref{cut1}).
The Borel-integration over $t > 0$ 
($\Leftrightarrow \ b > n \pm i \varepsilon$)
in Eq.~(\ref{cut1pp}) converges only when ${\rm Re}(\nu) < 0$ 
and gives the result (\ref{cut1ppp}).
When ${\rm Re}(\nu) \geq 0$, the result
(\ref{cut1ppp}) represents the analytic continuation in $\nu$.

The subtraction of the cut-function contribution (\ref{cut1}) 
leads to the final result for the resummed value
of the observable at the positive real value
$z = |z| = \beta_0 a(Q)/n > 0$
\begin{eqnarray}
\Delta[z\!=\!\beta_0 a(Q)/n] &=& \Delta_{\rm BI}(|z| \pm i \varepsilon) 
- \Delta_{\rm (cut1)}(|z| \pm i \varepsilon) =
{\tilde \Delta}_{\rm BI}[z\!=\!\beta_0 a(Q)/n] 
\nonumber \\
&=&
\left[ {\rm Re} \mp \cot(\pi \nu)~{\rm Im} \right]
\int_{\pm i \varepsilon}^{\infty\!\pm\!i \varepsilon} 
\frac{db}{\beta_0} \exp \left[- \frac{b}{\beta_0 a(Q)} \right]
\frac{R(b)}{(1 - b/n)^{1 + \nu}} .
\label{res1}
\end{eqnarray}
In the Borel integration here, the exactly known IR renormalon
singularity has been factored out explicitly. Function
$R(b)\!=\!(1 - b/n)^{1 +\nu} B(b)$ 
(whose truncated perturbation series up to $\sim\!b^2$ 
is known exactly in the case of several QCD observables)
has a much weaker
singularity at $b\!=\!n$ than $B(b)$.
In Eq.~(\ref{res1}), the equality
${\rm Re} \Delta_{\rm (cut1)}(|z| \pm i \varepsilon) =
\pm \cot(\pi \nu)~{\rm Im} \Delta_{\rm (cut1)}(|z| \pm i \varepsilon)$
was taken into account, a direct consequence of
Eqs.~(\ref{ImDBI}), (\ref{DBI}), and
 $(-|z|\!\mp\!i \varepsilon)^{-\nu}\!=\!
|z|^{-\nu} [ \cos(\pi \nu)\!\pm\!i \sin(\pi \nu)]$.
The results (\ref{cut1}) and (\ref{res1}) are those
arrived at in Ref.~\cite{Lee:2001ws}.

The renormalon power coefficient $\nu = (n c_1 - \gamma_{2n})/\beta_0$
should be regarded as a general parameter which
can take on, in principle, any (real) value.
For example, when varying the number of effectively
active quark flavors $n_f$ continuously,
$\nu$ changes continuously. Yet another example
is given by the large-$\beta_0$ approximation, 
when $c_1 \to 0$ and $\nu \to 0$. 
The residue $C$ of the renormalon in general does not vanish
in the large-$\beta_0$ limit. 

The cut-function (\ref{cut1}),
as a function of general $\nu$ and when the coupling
parameter is near the positive real axis $z_{\pm} =
|z| \pm i \varepsilon =
\beta_0 a(Q)/n \pm i \varepsilon$, can be rewritten
in the following form:
\begin{eqnarray}
\Delta_{\rm (cut1)}(z=|z| \pm i \varepsilon) &=& 
- C \frac{n}{\beta_0} ~e^{-1/|z|} 
\frac{ \left[ - \pi \cot( \pi \nu) 
\mp i \pi \right]}{\Gamma(1\!+\!\nu)}  
~|z|^{-\nu} \ .
\label{cut1b}
\end{eqnarray}
The central assumption of the method, i.e., the
subtraction of the cut-function (\ref{cut1}) from the
Borel-integrated value, appears to be plausible and
natural, especially because the cut function
(\ref{cut1}) [$\Leftrightarrow$ (\ref{cut1b})]
has a simple form, 
and because it represents precisely the contribution of the
Borel integration of the
nonanalytic part of the Borel transform (\ref{IR1})
along the cut $b \geq n$ parallel to the real axis, 
as explained in Eqs.~(\ref{cut1p})--(\ref{cut1ppp}).
However, there are at least two problems with
this method when we impose on it the plausible condition
that it should work for any (real) value of the
power coefficient $\nu$.
\begin{enumerate}
\item
When $\nu$ is nonnegative integer
($\nu = k;$ $k=0,1,2,\ldots$),
the cut-function (\ref{cut1b}) is infinite, because
$\cot(\pi \nu)$ diverges there. This would not present
a problem if the residue $C$ of the
renormalon disappeared, making this cut-function
finite. This appears to be
unlikely, as argued above; in particular, for
Adler function in the large-$\beta_0$
approximation ($n=2, c_1=0, \gamma_{4}=0$)
this would mean the disappearance of the
leading IR renormalon. Note that, in contrast to $\cot( \pi \nu)$,
the factor $1/\Gamma(1+\nu)$ in Eq.~(\ref{cut1b}) is an analytic
function in the entire complex $\nu$-plane.
\item
When $\nu$ is
a negative integer $\nu = -1, -2, \ldots,...$,
the IR renormalon singularity (\ref{IR1}) with the cut
disappears (even when $C \not= 0$), the Borel transform is analytic.
This means that 
$\Delta_{\rm (cut1)}(z)$
must be zero. However, according to Eq.~(\ref{cut1b}) 
$\Delta_{\rm (cut1)}(z) \not= 0$
-- note: 
$\pi \cot(\pi \nu)/\Gamma(1+\nu) = (-1)^{k} (k-1)!
\not=0$ when $\nu = -k = -1, -2, \ldots$. Again, it is
the poles of $\cot(\pi \nu)$ that cause the problems,
but this time at $\nu = -1, -2, \ldots$.
\end{enumerate} 

If we take the view-point that the method
should be taken as the starting point nonetheless,
because of the mentioned plausibility and naturalness of the
choice of (\ref{cut1}), we should definitely
modify it so that the aforementioned two 
problematic points are eliminated. 
This can be done in a natural way, by
inspecting again the expression (\ref{cut1b}).
The two problematic aspects arose because of the
poles of the term $\cot(\pi \nu)$, at
$\nu = 0, \pm1, \pm2$, etc. 
The function $f(\nu) = \cot(\pi \nu)$ should thus
be regularized in order to eliminate both
problems. The apparently most natural regularization
is obtained by subtracting the simple pole functions
with the coefficients being equal to the residues of $f(\nu)$
at those poles. This gives:
\begin{eqnarray}
- \pi \cot(\pi \nu) &\mapsto& - \pi \cot(\pi \nu) + \frac{1}{\nu}
+ \sum_{k=1}^{\infty} \left( \frac{1}{\nu -k} + \frac{1}{\nu + k}
\right)  
\label{sub1}
\\
& = & - \pi \cot(\pi \nu) + \frac{1}{\nu} + 
2 \nu \sum_{k=1}^{\infty} \frac{1}{\nu^2 - k^2} \equiv 0
\label{sub2}
\end{eqnarray}
We see that the function $\cot(\pi \nu)$ is identical to
the sum of the corresponding simple pole functions. Only
some of the meromorphic functions
have this remarkable property.
The procedure (\ref{sub1})--(\ref{sub2})
therefore eliminates the real part of the cut-function
(\ref{cut1b}) at the real axis $z = |z|\!\pm\!i \varepsilon$,
the modified expression there is purely imaginary.
The final result of the modified method is then
just the Principal Value (PV) of the Borel integral, when $a(Q) >0$
\begin{eqnarray}
\Delta[z=\beta_0 a(Q)/n] &=& {\rm Re} \Delta_{\rm BI}(z \pm i \varepsilon) 
= {\rm Re}~\int_{\pm i \varepsilon}^{\infty\!\pm\!i \varepsilon} 
\frac{db}{\beta_0} \exp \left[- \frac{b}{\beta_0 a(Q)} \right]
\frac{R(b)}{(1 - b/n)^{1 + \nu}} 
\ ,
\label{PV1}
\end{eqnarray}
and the new cut-function can be written for
a general complex $z$ as
\begin{eqnarray}
\Delta^{\rm(PV)}_{\rm cut}(z) & = & + C \frac{n}{\beta_0}~e^{-1/z}
~\frac{\pi}{\Gamma(1\!+\!\nu)~\sin(\pi \nu)}
\left[ (-z)^{-\nu} - \cos(\pi \nu)~z^{-\nu} \right]
\ .
\label{cutPVplane}
\end{eqnarray}
It can be checked explicitly that this function is
finite for any finite complex $\nu$, including
$\nu = 0, \pm 1, \pm 2, \ldots$. It is an analytic
function of $z$ outside the real axis, 
with the cut along the real axis.
The subtraction of the poles of $\cot(\pi \nu)$ (\ref{sub2})
introduced in (\ref{cutPVplane}) terms proportional
to $z^{-\nu}$. The consistency
of the results for $\nu = 0, \pm1, \ldots$ thus
introduced an additional cut along the negative real axis.

In general, any regularization in $\nu$ of the singular
$\cot(\pi \nu)$ factor in Eq.~(\ref{cut1b}), not just 
the simplest regularization (\ref{sub1}), would give
an acceptable result. The factor $1/\Gamma(1\!+\!\nu)$
is already nonsingular (even analytic) for all $\nu$.
The factor $\exp(-1/|z|) |z|^{-\nu}$ in Eq.~(\ref{cut1b})
cannot be modified because it reflects the correct
$Q$-dependence of the aforementioned higher-twist
term $\langle {\cal O}_{2 n} \rangle^{(Q)}/ (Q^2)^n$
in the OPE. The imaginary part in Eq.~(\ref{cut1b})
cannot be modified because it is needed to make the amplitude 
$\Delta(z) \equiv \Delta_{\rm BI}(z)\!-\!\Delta_{\rm cut}(z)$
real for the real positive $z$.
Therefore, the most general $\nu$-regularization of
the cut-function (\ref{cut1b}) for $z = |z|\!\pm\!i \varepsilon$
is represented by a simple substitution 
$\cot (\pi \nu) \mapsto g(\nu)$, where $g(\nu)$
is an arbitrary nonsingular (possibly analytic)
function of $\nu$ which is real for real $\nu$
\begin{equation}
\Delta_{\rm cut}(z\!=\!|z|\!\pm\!i \varepsilon)
= - C \frac{n}{\beta_0} ~e^{-1/|z|} 
\frac{ \left[ - \pi g(\nu) \mp i \pi \right]}{\Gamma(1\!+\!\nu)}  
~|z|^{-\nu} \ .
\label{cut1c}
\end{equation}
This cut function has then the following form in the
complex coupling plane:
\begin{equation}
\Delta_{\rm cut}(z)  =  + C \frac{n}{\beta_0}~e^{-1/z}
~\frac{\pi}{\Gamma(1\!+\!\nu)~\sin(\pi \nu)}
\left\{ (-z)^{-\nu} - \left[ \cos(\pi \nu) - g(\nu) \sin(\pi \nu)
\right]~z^{-\nu} \right\}
\ .
\label{cutplane}
\end{equation}
Again, as in Eq.~(\ref{cutPVplane}),
we see that the $\nu$-regularization introduces 
an additional cut along the negative axis $z < 0$.
The amplitude $\Delta\equiv \Delta_{\rm BI}\!-\!\Delta_{\rm cut}$
is now represented by the Principal Value and a higher-twist
term proportional to $g(\nu)$ where $g(\nu)$ is nonsingular
in $\nu$. We stress that any such choice is physically
acceptable, the Principal Value choice being distinguished
in this context only by the mathematical simplicity
of the corresponding $\nu$-regularization
(\ref{sub1})--(\ref{sub2}).

The main idea behind the result Eqs.~(\ref{cut1}), (\ref{res1}),
as stressed in Ref.~\cite{Lee:2001ws},  was that the
cut-function which is to be subtracted from a Borel-resummed
QCD amplitude with an IR renormalon has a cut only along the
positive axis in the coupling plane $z \equiv a(Q)$.
In the two exactly solvable non-QCD examples presented in
Ref.~\cite{Lee:2001ws} this idea was shown to hold.
Here we showed that the requirement of finiteness with
respect to the IR renormalon power parameter $\nu$
in QCD amplitudes, and their consistency in the absence 
of the IR renormalon, imply that the cut function
must contain, in addition to the cut along the positive
$z$-axis, also a cut along the negative $z$-axis.
This, in turn, implies that the amplitude
$\Delta(z) \equiv \Delta_{\rm BI}(z)\!-\!\Delta_{\rm cut}(z)$,
while having no cut on the $z > 0$ axis in accordance with
the unitarity and causality conditions \cite{Oehme:1994pv},
does have a cut along the $z < 0$ axis (Landau region)
as a consequence of the IR renormalon.\footnote{
According to Ref.~\cite{Oehme:1994pv}, the physical
singularities in confined theories are generated
by the physical hadron states. The point $z=0$ is
an essential singularity \cite{'tHooft:am,Khuri:xc}.
The confinement is not seen by pQCD, thus
the cut along $z < 0$ (Landau region) is not physical
but must appear in quantities involving pQCD coupling $a(Q)$.}
This somewhat counterintuitive conclusion was
also obtained in Refs.~\cite{Caprini:1999ma,Caprini:2001mn}
by substantially different approaches, where the
Principal Value prescription was adopted. 
If the considered amplitude has ultraviolet (UV)
renormalons, it should have a cut along $z < 0$
even if it has no IR renormalons.
If the coupling $a(Q)$ is regularized 
so that it is finite for all $Q^2 \geq 0$ (see, e.g.,
\cite{ECH,Mattingly:ej,csr,AQCD,Dokshitzer:1995qm}),
i.e., in contrast to the pQCD $a(Q)$ it has no Landau 
singularities, the conclusions about the analyticity
of the considered amplitude in the complex coupling plane 
probably change singnificantly.

The IR renormalons play an important role also
in the resummations using modified Borel transforms
where the entire integrand in the Borel integration
is renormalization scale (RS) invariant. Such transforms
were introduced by Grunberg \cite{Grunberg:1993hf} on the basis of
a larger class of transforms proposed in Ref.~\cite{modBT}
in a somewhat different context.
Such RS-invariant Borel transform resummations were applied 
in Refs.~\cite{Cvetic:2001ws,Cvetic:2001pg}, by either
evaluating the Principal Value of the Borel integral \cite{Cvetic:2001ws} 
or adding to the Principal Value the
higher-twist OPE terms \cite{Cvetic:2001pg}.
The discussed method of Ref.~\cite{Lee:2001ws}, for the ordinary
Borel transforms (\ref{Bintegraldef}), can be
adapted to the method of the RS-invariant
Borel transforms. The problems (divergences) appearing
in this case are similar to those discussed here,
but algebraically more complicated. 
It is not clear whether in such case an analogous
regularization procedure as the one presented here
would lead naturally to the Principal Value of the RS-invariant
Borel resummation.

In the present work, the method of
subtracting a power-suppressed term from the
Principal Value of the Borel integral
for QCD amplitudes with IR renormalon, recently proposed 
in Ref.~\cite{Lee:2001ws} and applied in 
Refs.~\cite{Contreras:2002kf},
was scrutinized. It was pointed out that 
the result becomes physically
untenable for specific values of the renormalon
power coefficient $\nu$, as a consequence of
the divergences of a term [$\cot (\pi \nu)$]
appearing in the result. When these
divergences are removed in apparently the most natural way,
the power-suppressed term of the method disappears
and the modified result becomes the Principal Value.
Any removal of the aforementioned divergences results
in an IR-renormalon-induced cut along the negative axis
(Landau region) in the coupling plane. 
These conclusions suggest, among other things, that the most
natural pQCD Borel integration of a QCD observable
remains the Principal Value.
The additional power-suppressed
(higher-twist, higher-dimensional) terms
cannot be inferred from pQCD (+renormalon)
methods in any natural way. Such additional (OPE)
terms involve vacuum expectation values of
higher-twist operators and can theoretically
be obtained or estimated only by genuinely
nonperturbative methods. 
Phenomenologically, such additional OPE terms can
be determined by fitting them to the corresponding
experimental data. However, in such a procedure,
it is important to keep for the leading-twist term
in the OPE a specific resummed pQCD expression,
most naturally the Principal Value
of the Borel integral, i.e., Eq.~(\ref{PV1}).
On the other hand, if the leading-twist term is taken to be
a truncated perturbation series (TPS), the strength
of the higher-twist terms will sometimes
dramatically change when the order of the
TPS is changed \cite{KPS,yang}.

\begin{acknowledgments}
This work was supported by FONDECYT (Chile) 
grant No. 1010094.
\end{acknowledgments}


\begin{thebibliography}{99}

\bibitem{Beneke:1999ui}
M.~Beneke,
%``Renormalons,''
Phys. Rept. {\bf 317}, 1 (1999), and references therein.
%%CITATION = PRPLC,317,1;%%

%\cite{Khuri:bf}
\bibitem{Khuri:bf}
N.~N.~Khuri,
%``Zeros Of The Gell-Mann-Low Function And Borel Summations 
%In Renormalizable Theories,''
Phys.\ Lett.\ B {\bf 82}, 83 (1979);
%%CITATION = PHLTA,B82,83;%%
%\bibitem{Lovett-Turner:1995ti}
C.~N.~Lovett-Turner and C.~J.~Maxwell,
%``All orders renormalon resummations for some QCD observables,''
Nucl.\ Phys.\ B {\bf 452}, 188 (1995)
[arXiv:hep-ph/9505224];
%%CITATION = HEP-PH 9505224;%%
%\bibitem{Ball:1995ni}
P.~Ball, M.~Beneke and V.~M.~Braun,
%``Resummation of (beta0 alpha-s)**n corrections in QCD: 
%Techniques and applications to the tau hadronic width and 
%the heavy quark pole mass,''
Nucl.\ Phys.\ B {\bf 452}, 563 (1995)
[arXiv:hep-ph/9502300];
%%CITATION = HEP-PH 9502300;%%
%\bibitem{Caprini:1998wg}
I.~Caprini and J.~Fischer,
%``Accelerated convergence of perturbative {QCD} by 
%optimal conformal  mapping of the Borel plane,''
Phys.\ Rev.\ D {\bf 60}, 054014 (1999)
[arXiv:hep-ph/9811367].
%%CITATION = HEP-PH 9811367;%%

%\cite{Cvetic:2001sn}
\bibitem{Cvetic:2001sn}
G.~Cveti\v{c} and T.~Lee,
%``Bilocal expansion of Borel amplitude and hadronic tau decay width,''
Phys.\ Rev.\ D {\bf 64}, 014030 (2001)
[arXiv:hep-ph/0101297].
%%CITATION = HEP-PH 0101297;%%

%\cite{Caprini:1999ma}
\bibitem{Caprini:1999ma}
I.~Caprini and M.~Neubert,
%``Borel summation and momentum-plane analyticity in perturbative {QCD},''
JHEP {\bf 9903}, 007 (1999)
[arXiv:hep-ph/9902244].
%%CITATION = HEP-PH 9902244;%%

%\cite{Caprini:2001mn}
\bibitem{Caprini:2001mn}
I.~Caprini and J.~Fischer,
%``Analytic continuation and perturbative expansions in QCD,''
Eur.\ Phys.\ J.\ C {\bf 24}, 127 (2002)
[arXiv:hep-ph/0110344].
%%CITATION = HEP-PH 0110344;%%

%\cite{Lee:2001ws}
\bibitem{Lee:2001ws}
T.~Lee,
%``Nonperturbative effects from the resummation of perturbation theory,''
Phys.\ Rev.\ D {\bf 66}, 034027 (2002)
[arXiv:hep-ph/0104306].
%%CITATION = HEP-PH 0104306;%%

%\cite{Contreras:2002kf}
\bibitem{Contreras:2002kf}
C.~Contreras, G.~Cveti\v{c}, K.~S.~Jeong and T.~Lee,
%``Extraction of alpha(s) from the Gross-Llewellyn Smith 
%sum rule using  Borel resummation,''
Phys.\ Rev.\ D {\bf 66}, 054006 (2002)
[arXiv:hep-ph/0203201];
%%CITATION = HEP-PH 0203201;%%
%\bibitem{Lee:2002sn}
T.~Lee,
%``Surviving the renormalon in heavy quark potential,''
arXiv:hep-ph/0210032.
%%CITATION = HEP-PH 0210032;%%

%\cite{GR}
\bibitem{GR}
I.~S.~Gradshteyn and I.~M.~Ryzhik, Table of integrals, series
and products (Academic Press, Inc., 1980).
%%CITATION = NONE;%%

%\cite{Oehme:1994pv}
\bibitem{Oehme:1994pv}
R.~Oehme,
%``Analytic structure of amplitudes in gauge theories with confinement,''
Int.\ J.\ Mod.\ Phys.\ A {\bf 10}, 1995 (1995)
[arXiv:hep-th/9412040];
%%CITATION = HEP-TH 9412040;%%
%\bibitem{Oehme:yr}
%``On The Analytic Structure Of Hadronic Amplitudes In QCD,''
arXiv:hep-ph/9212292.
%%CITATION = HEP-PH 9212292;%%

%\cite{'tHooft:am}
\bibitem{'tHooft:am}
G.~'t Hooft, in: {\em The Whys of Subnuclear Physics,}
Proceedings of the 15th International School on Subnuclear Physics, 
p.~943, Erice, Sicily, 1977, A.~Zichichi ed. 
(Plenum Press, New York 1979).
%%CITATION = NONE;%%

%\cite{Khuri:xc}
\bibitem{Khuri:xc}
N.~N.~Khuri,
%``Coupling Constant Analyticity And The Renormalization Group,''
Phys.\ Rev.\ D {\bf 23}, 2285 (1981).
%%CITATION = PHRVA,D23,2285;%%

\bibitem{ECH}
G.~Grunberg, Phys. Lett. {\bf 95B}, 70 (1980),
{\bf 110B}, 501(E) (1982);
%%CITATION = PHLTA,B95,70;%%
{\bf 114B}, 271 (1982);
%%CITATION = PHLTA,B114,271;%%
Phys. Rev. D {\bf 29}, 2315 (1984);
%%CITATION = PHRVA,D29,2315;%% 
A.~L.~Kataev, N.~V. Krasnikov, and A.~A. Pivovarov,
Nucl. Phys. {\bf B198}, 508 (1982);
%%CITATION = NUPHA,B198,508;%% 
A.~Dhar and V.~Gupta, Phys. Rev. D {\bf 29}, 2822 (1984);
%%CITATION = PHRVA,D29,2822;%%
V.~Gupta, D. V. Shirkov, and O. V. Tarasov, 
Int. J. Mod. Phys. A {\bf 6}, 3381 (1991).
%%CITATION = IMPAE,A6,3381;%%

%\cite{Mattingly:ej}
\bibitem{Mattingly:ej}
A.~C.~Mattingly and P.~M.~Stevenson,
%``Optimization Of R(E+ E-) And 'Freezing' Of The QCD 
%Couplant At Low-Energies,''
Phys.\ Rev.\ D {\bf 49}, 437 (1994)
[arXiv:hep-ph/9307266].
%%CITATION = HEP-PH 9307266;%%

\bibitem{csr}
H.~J.~Lu and S.~J.~Brodsky, Phys. Rev. D {\bf 48}, 3310 (1993);
%%CITATION = PHRVA,D48,3310;%% 
S. J. Brodsky and H. J. Lu, {\em ibid.\/} {\bf 51}, 3652 (1995);
%%CITATION = PHRVA,D51,3652;%% 
%\bibitem{Brodsky:2002nb}
S.~J.~Brodsky, S.~Menke, C.~Merino and J.~Rathsman,
%``On the behavior of the effective QCD coupling alpha(tau)(s) 
%at low  scales,''
arXiv:hep-ph/0212078.
%%CITATION = HEP-PH 0212078;%%

\bibitem{AQCD}
D.~V.~Shirkov and I.~L.~Solovtsov, 
Phys. Rev. Lett. {\bf 79}, 1209 (1997);
%%CITATION = PRLTA,79,1209;%% 
K. A. Milton, I. L. Solovtsov, and O. P. Solovtsova,
Phys. Lett. B {\bf 439}, 421 (1998);
%%CITATION = PHLTA,B439,421;%% 
D.~V.~Shirkov, hep--ph/0003242.
%%CITATION = HEP-PH 0003242;%%

%\cite{Dokshitzer:1995qm}
\bibitem{Dokshitzer:1995qm}
Y.~L.~Dokshitzer, G.~Marchesini and B.~R.~Webber,
%``Dispersive Approach to Power-Behaved Contributions in QCD Hard Processes,''
Nucl.\ Phys.\ B {\bf 469}, 93 (1996)
[arXiv:hep-ph/9512336].
%%CITATION = HEP-PH 9512336;%%

%\cite{Grunberg:1993hf}
\bibitem{Grunberg:1993hf}
G.~Grunberg,
%``The Renormalization scheme invariant Borel transform and 
%the QED renormalons,''
Phys.\ Lett.\ B {\bf 304}, 183 (1993).
%%CITATION = PHLTA,B304,183;%%

\bibitem{modBT}
%\cite{Brown:1992ic}
%\bibitem{Brown:1992ic}
L.~S.~Brown and L.~G.~Yaffe,
%``Asymptotic behavior of perturbation theory for 
%the electromagnetic current current correlation function in QCD,''
Phys.\ Rev.\ D {\bf 45}, 398 (1992);
%%CITATION = PHRVA,D45,398;%%
%\cite{Brown:1992pk}
%\bibitem{Brown:1992pk}
L.~S.~Brown, L.~G.~Yaffe and C.~Zhai,
%``Large order perturbation theory for the 
%electromagnetic current-current correlation function,''
%Phys.\ Rev.\ D {\bf 46}, 4712 (1992)
{\em ibid.\/} {\bf 46}, 4712 (1992)
[hep-ph/9205213].
%%CITATION = HEP-PH 9205213;%%

%\cite{Cvetic:2001ws}
\bibitem{Cvetic:2001ws}
G.~Cveti\v{c}, C.~Dib, T.~Lee and I.~Schmidt,
%``Resummation of the hadronic tau decay width 
%with modified Borel  transform method,''
Phys.\ Rev.\ D {\bf 64}, 093016 (2001)
[arXiv:hep-ph/0106024];
%%CITATION = HEP-PH 0106024;%%

%\cite{Cvetic:2001pg}
\bibitem{Cvetic:2001pg}
G.~Cveti\v{c}, T.~Lee and I.~Schmidt,
%``Resummations with renormalon effects for 
%the leading hadronic  contribution to the muon (g-2),''
Phys.\ Lett.\ B {\bf 520}, 222 (2001)
[arXiv:hep-ph/0107069].
%%CITATION = HEP-PH 0107069;%%

\bibitem{KPS}
%\bibitem{Kataev:1997nc}
A.~L.~Kataev, A.~V.~Kotikov, G.~Parente and A.~V.~Sidorov,
%``Next-to-next-to-leading order QCD analysis of the revised CCFR data 
%for  xF3 structure function,''
Phys.\ Lett.\ B {\bf 417}, 374 (1998)
[arXiv:hep-ph/9706534];
%%CITATION = HEP-PH 9706534;%%
%\bibitem{Kataev:1999bp}
A.~L.~Kataev, G.~Parente and A.~V.~Sidorov,
%``Higher twists and alpha(s)(M(Z)) extractions from the NNLO {QCD} 
%analysis  of the CCFR data for the xF3 structure function,''
Nucl.\ Phys.\ B {\bf 573}, 405 (2000)
[arXiv:hep-ph/9905310];
%%CITATION = HEP-PH 9905310;%%
%\bibitem{Kataev:2001kk}
A.~L.~Kataev, G.~Parente and A.~V.~Sidorov,
%``Fixation of theoretical ambiguities in the improved fits to 
%the xF3  CCFR data at the next-to-next-to-leading order and beyond,''
arXiv:hep-ph/0106221.
%%CITATION = HEP-PH 0106221;%%

\bibitem{yang}
U.~K.~Yang and A.~Bodek,
%``Parton distributions, d/u, and higher twist effects at high x,''
Phys.\ Rev.\ Lett.\  {\bf 82}, 2467 (1999)
[arXiv:hep-ph/9809480];
%%CITATION = HEP-PH 9809480;%%
%\bibitem{Yang:1998nj}
U.~K.~Yang, A.~Bodek and Q.~Fan,
%``Parton distributions, d/u, and higher twists at high x,''
arXiv:hep-ph/9806457.
%%CITATION = HEP-PH 9806457;%%

\end{thebibliography}
\end{document}